\documentclass[preprint,12pt]{elsarticle}

\usepackage{graphics}
\usepackage{graphicx}

\usepackage{amsmath}
\usepackage{amsthm}

\biboptions{comma,sort,compress}

\journal{Physica A}

\begin{document}

\begin{frontmatter}



\title{Modeling of income distribution in the European Union with the Fokker--Planck equation}


\author{Maciej Jagielski\corref{cor1}}
\ead{zagielski@interia.pl}
\cortext[cor1]{Corresponding author. Tel.: +48 22 5533229; fax: +48 22 5532410.}
\author{Ryszard Kutner}%

\address{Institute of Experimental Physics\\
Faculty of Physics, University of Warsaw\\
Ho\.za 69, PL-00681 Warszawa, Poland}

\begin{abstract}
Herein, we applied statistical physics to study incomes of three (low-, medium- and high-income) society classes instead of the two (low- and medium-income) classes studied 
so far. In the frame of the threshold nonlinear Langevin dynamics and its threshold Fokker-Planck counterpart, 
we derived a unified formula for description of income of all  society classes, by way of example, of those of the European 
Union in year 2006 and 2008. Hence, the formula is more general than the well known
that of Yakovenko et al. That is, our formula well describes not only two
regions but simultaneously the third region in the plot of the complementary
cumulative distribution function vs. an annual household income. Furthermore,
the known stylised facts concerning this income are well described by our
formula. Namely, the formula provides the Boltzmann-Gibbs income distribution function
for the low-income society class and 
the weak Pareto law for the medium-income society class, as expected. Importantly, it predicts (to satisfactory 
approximation) the Zipf law for the high-income society class. Moreover, the region of medium-income 
society class is now distinctly reduced because the bottom of high-income society class is distinctly lowered. This 
reduction made, in fact, the medium-income society class an intermediate-income society class.  
\end{abstract}


\end{frontmatter}


\section{Introduction}
For over two decades, physics oriented approaches have widely been developed to explain different economic processes
\cite{Yako1, Yako2, RS, SR, RHCR, Drozdz, Stan} (and refs. therein). Those approaches aim at formulating well fitted 
unbiased indicators of social and economic phenomena. One of their key issues is the income of society analysis using methods of statistical physics. The main goal of this economic issue is to unravel and describe mechanisms 
of societies' enrichment or impoverishment.

The first successful attempt in this socio-economic field was made by the legendary economist and sociologist 
Vilfredo Pareto \cite{RHCR, Mandel, Mandel1}. He demonstrated that the distribution functions of individual incomes in 
different countries within stable economy are universal being manifested by a power law. This law is called the weak 
Pareto law.  
He emphasised that this law could not resemble the distribution functions obtained if the gain and accumulation 
of income were random. As a possible origin of this law, Pareto indicated a self-similarity structure of societies.

Pareto's economic discoveries initiated attempts of analytical descriptions of incomes of the societies and inspired 
an avalanche of related research works 
\cite{Yako1, Yako2, RS, SR, RHCR, Arma, Suto, Champ, ChCh, BChCh, DY, DY1, CMGK, FMG, BYM, MAH, FM, PLAO, ME, ME1}. 
Among them, particularly significant are those of the economist Robert Gibrat \cite{RHCR, Arma, Suto}. He found that 
the complementary cumulative distribution function of the Pareto distribution is insufficient to describe empirical 
data within the whole range of the income. Trying to find a functional form that could account for these data, he 
proposed a rule called the Rule of Proportionate Growth \cite{Arma, Suto} (see below for details).

Furthermore, the income of societies was analysed by David Champerowne, who constructed a stochastic model simulating 
the Pareto power law \cite{Champ} and also by Benoit Mandelbrot who described several useful properties of random 
variables subjected the Pareto distribution \cite{Mandel, Mandel1}. 

In the recent decade, a large number of studies were performed aiming at constructing of models, which 
(to some extend) would well  replicate the observed complementary cumulative distribution functions of individual 
incomes. Among them, the most significant seems to be the Clementi-Matteo-Gallegati-Kaniadakis approach \cite{CMGK}, 
the Generalized Lotka-Volterra Model \cite{SR, RS, RHCR}, the Boltzmann-Gibbs law \cite{ChCh, BChCh, DY,DY1}, and the 
Yakovenko et al. model \cite{Yako1, Yako2}. Very recently, a mathematical model similar to that of Yakovenko et al. 
has been developed \cite{FM}. It involves complex economic justification for microscopic stochastic dynamics of wealth. 
However, none of the above attempts to find an analytical description of the income structure solves the principal 
challenges, which concern:
\begin{itemize}
\item[(i)] the description of the annual household incomes of all society classes (i.e. the low-, medium-, and high-income society classes) by a single unified formula and
\item[(ii)] the problem regarding corresponding complete microscopic (microeconomic) mechanism responsible for 
the income structure and dynamics. 

In our considerations presented herein, we used Boltzmann-Gibbs law and Yakovenko et al. model to derive a uniform
analytical formula describing income of all three society classes.
\end{itemize}

\section{Extended Yakovenko et al. model}
In accord with an effort outlined above, we compared the empirical data of the annual household incomes in the European 
Union (EU), including Norway and Iceland, with predictions of our theoretical approach proposed herein. This approach 
is directly inspired by the Yakovenko et al. model. By using the same assumptions, however, we generalised this model 
to solve the principal challenges (i) and (ii) indicated above.

We used data records from the Eurostat Survey on Income and Living Conditions
(EU-SILC) \cite{EURO3}, by way of example for years 2006 and 2008 \cite{EURO1,
EURO2} (containing around $150$ and $200$ thousand empirical data points,
respectively). However, these records contain (as all other records) only few
data points concerning the high-income society class, i.e. the third region in
the plot of the complementary cumulative probability distribution function vs.
annual household income. To consider the high-income society class systematically, we additionally analysed 
the effective income of  billionaires in the EU by using the Forbes 'The World's Billionaires' rank \cite{Forbes}. 
The term 'billionaire' used herein is equivalent (as in  the US terminology) to the term 'multimillionaire' used 
in the European terminology. Since we consider wealth and income of billionaires in euros, we recalculated US dollars 
to euros by using the mean exchange rate at the day of construction of the Forbes 'The World's Billionaires' rank.   

We were able to consider incomes of  three society classes  thanks to  the following  procedure.
\begin{itemize}
\item[(i)] Firstly, we selected EU billionaires' wealth from the Forbes 'The World's Billionaires' rank, for instance, 
for four successive years  2005 to 2008.
\item[(ii)] Secondly, we calculated the corresponding differences between billionaires' wealth for the successive years.
We assumed that their incomes are, in fact, proportional to these differences. For instance, we calculated the 
billionaire incomes for year 2006 by taking the difference between their wealth in years 2006 and 2005. We made it 
analogously for year 2008. However, we took into account only billionaires who gained effective incomes (neglecting 
those, who suffered from income losses).
\item[(iii)] Subsequently, having so calculated incomes for the high-income society class, we joined them (separately 
for years 2006 and 2008) with the corresponding EU-SILC datasets. By using so completed datasets, we then constructed 
the initial empirical complementary cumulative distribution function for years 2006 and 2008. For that, we used the well known 
Weibull recipe (see below for details). However, this direct 
approach shows a wide gap of incomes inside the high-income society class resulting in a horizontal line of the 
complementary cumulative distribution function. This gap separates the first segment belonging to the high-income 
society class, consisting of all data points taken from the EU-SILC dataset (only 8 for 2006 and 6 for 2008), from the 
second segment, consisting of remaining data points (76 for 2006 and 96 for 2008), which also belong to the high-income 
society class but are taken from the Forbes dataset.
\item[(iv)] In the final step, we eliminated this gap by adopting the assumption that the empirical 
complementary cumulative distribution functions (concerning the whole society) have no horizontal segments. That is, 
we assumed that statistics  of incomes is a continuous function of income. Hence, we were forced to multiply the billionaire
incomes from Forbes dataset by the properly chosen common proportionality factor. This factor was equal to 
$1.0\times 10^{-2}$ for both years, as we assumed the requirement of full overlap of the first (above mentioned) segment
by the second segment. This assumption leads to a unique solution (up to some negligible statistical error) for this 
proportionality factor. We found that this factor was only a slowly-varying function of time (or years).
\end{itemize}

Hence, we received the data record containing already a sufficient number of data points for all society classes, including 
the high-income society class. Although the Forbes empirical data only roughly estimate the wealth of billionaires, 
they quite well establish the billionaires' rank, thus sufficiently justifying our approach. This is because our 
purpose is to classify billionaires to concrete universality class rather than finding their total incomes. Our 
procedure of linking data from two different bases does not violate this universality class. 

The basic tool of our analysis is an empirical complementary cumulative distribution function being typical in this 
context. We calculated it according to the standard two-step procedure. For that, first, the income empirical data 
were ordered according to their rank, i.e. from incomes of the richest households to those of the poorest. Next, in 
accordance with the well known Weibull formula \cite{WCH, Chow}, we calculated the ratio 
$\frac{l}{n+1}$ where $l$ is the position of the household in the rank and $n$ is the size of the empirical data 
record. This ratio directly determines the required fraction of households of the income higher than that 
related to a given household position $l$ in the rank. The complementary cumulative distribution function obtained 
that way is sufficiently stable. Furthermore, it does not reduce the size of the output compared to that of the 
original empirical data record.

\subsection{Hint to the Yakovenko et al. model}
Let $m$ be an influx of income per unit time to a given household. We treat $m$ as a variable obeying stochastic 
dynamics. Then, we can describe its time evolution by using the nonlinear Langevin equation \cite{Yako1, Yako2, Kamp}:  
\begin{eqnarray}
\frac{dm}{dt}=-A(m)+C(m)\, \eta(t).
\label{rown1}
\end{eqnarray}
Here, $A(m)$ is a drift term and $\eta(t)$ is a white noise, where the coefficient $C(m)$ is its 
$m$-dependent amplitude. As we prove below, already white noise is  herein sufficient to produce two different 
power-laws. Obviously, generalisation with respect to power-law noise is also possible and very promising \cite{Strok1,Strok2}. 
Noteworthly, jump effects are important in the financial and social phenomena because they naturally
produce power-law tails. However, correspondence between It\^o and Stratonovich representations is then no 
longer trivial. 

Notably, the above nonlinear stochastic dynamics equation is equivalent to the following 
Fokker-Planck (continuity) equation for the probability distribution function \cite{Kamp}:
\begin{eqnarray}
\frac{\partial}{\partial t}P(m,t)&=&-\frac{\partial}{\partial m}J(m,t), 
\label{rown2a}
\end{eqnarray}
where the flux density of probability (in the It\^o representation \cite{Kamp}) is
\begin{eqnarray}
J(m,t)&=&-A(m)P(m,t)-\frac{{\partial}}{\partial m}\left[B(m)P(m,t)\right].\nonumber \\
\end{eqnarray}
Here, $B(m)=C^2(m)/2$ while $P(m,t)$ is the temporal income distribution function. In general, functions 
$A(m)$ and $B(m)$ can be additionally determined by the first and second moment of the income change per unit time, 
respectively, only if these moments exist.

The equilibrium solution of Eq. (\ref{rown2a}), $P_{\rm eq}$, defined by vanishing of $J(m,t)$ \cite{Kamp}, takes the 
form:
\begin{eqnarray}
P_{\rm eq}(m)=\frac{const}{B(m)}\exp\left(-\int_{{m_{\rm init}}}^m\frac{A(m')}{B(m')}\, dm'\right)
\label{rown5}
\end{eqnarray}
where $m_{\rm init}$ is the lowest household income and $const$ is a normalisation factor. Fortunately, both It\^o 
and Stratonovitch representations \cite{Kamp} give almost the same equilibrium distribution function. These representations differ only by some preexponential factor.

Following the Yakovenko et al. model \cite{Yako1, Yako2}, we can assume that changes of income of the low-income 
society class are independent of the previous income gained. This assumption is justified because the income of 
households belonging to this class mainly takes the form of wages and salaries. The stochastic process associated 
with the mechanism of this kind is called the additive stochastic process. In this case, coefficients $A(m)$ and 
$B(m)$ take, obviously, the form of positive constants
\begin{eqnarray}
A(m)=A_0,\quad B(m)=B_0.
\label{rown6}
\end{eqnarray}
This choice of coefficients leads to the Boltzmann-Gibbs law with the exponential complementary cumulative 
distribution function \cite{Yako1, Yako2, ChCh, BChCh, DY, DY1}:
\begin{eqnarray}
\Pi(m) = \int_{m}^{\infty} P_{\rm eq}(m')\, dm' = \exp\left(-\frac{m-m_{\rm init}}{T}\right). 
\label{rown7}
\end{eqnarray}
In Equation (\ref{rown7}), distribution function is characterised by a single parameter, i.e. an income temperature 
$T=B_0/A_0$, which can be interpreted in this case as an average income per household.

For the medium- and high-income society classes, we can assume (again following Yakovenko et al. \cite{Yako1, Yako2}) 
that changes of income are proportional to the income gained so far. This assumption is also justified because profits 
go to the medium- and high-income society classes mainly through investments and capital gains. This type of stochastic 
process is called the multiplicative stochastic process. Hence, coefficients $A(m)$ and $B(m)$ obey the proportionality 
principle of Gibrat 
\cite{Arma, Suto}:
\begin{eqnarray}
A(m)=a\, m,\quad B(m)=b\, m^2\Leftrightarrow C(m)=\sqrt{2\, b}\, m,
\label{rown8}
\end{eqnarray}
where $a$ and $b$ are positive parameters. By using the equilibrium distribution function (\ref{rown5}), we arrive in this 
case to the weak Pareto law with the complementary cumulative distribution function \cite{Yako1, Yako2, RHCR}:
\begin{eqnarray}
\Pi(m) = \int_{m}^{\infty} P_{\rm eq}(m')\, dm' = \left(\frac{m}{m_{\rm sp}}\right)^{-\alpha}.
\label{rown9}
\end{eqnarray}
Here, $m_{\rm sp}$ is a scaling factor (depending on $a,\, b$, and $const$) while $\alpha=1+a/b$ is the Pareto exponent. The ratio of the $a$ to $b$ parameters can directly be determined from the empirical data expressed in the log-log plot (by using their slopes). 

As Yakovenko et al. have already found \cite{Yako1, Yako2}, the coexistence of additive and multiplicative stochastic 
processes is allowed. By assuming that these processes are uncorrelated, we get
\begin{eqnarray}
A(m)&=&A_0+am, \nonumber \\
B(m)&=&B_0+b\, m^2=b\, (m^2_0+m^2),  
\label{rown10}
\end{eqnarray}
where $m^2_0=B_0/b$. This consideration leads (together with Eq. (\ref{rown5})) to a significant Yakovenko et al. model 
with the probability distribution function given by
\begin{eqnarray}
P_{\rm eq}(m) = const\, \frac{e^{-(m_0/T)\arctan(m/m_0)}}{[1+(m/m_0)^2]^{(\alpha +1)/2}}
\label{rown11}
\end{eqnarray} 
where parameters $\alpha$ and $T$ are defined above. For $m\ll m_0$, Eq. (\ref{rown11})  becomes the 
Boltzmann-Gibbs law while for $m\gg m_0$ it becomes the weak Pareto law. Notably, the $m_0$ parameter is the crossover 
income between ranges of additive and multiplicative processes.

\subsection{Our extension}
Based on the Yakovenko et al. Eq. (\ref{rown11}), the complementary cumulative distribution function can be used to describe 
income of only low- and medium-income society classes. However, it does not capture that of the most intriguing 
high-income society class. Therefore, the goal of our present work is to derive from Eq. (\ref{rown5}) such a 
distribution function, which would cover all three ranges of the empirical data records, i.e. low-, medium-, and 
high-income classes of the society (including also two short intermediate regions between them).

The high-income society class is mainly that composed of the company owners. Hence, besides the weak Pareto law, 
we expect that their household incomes obey (to a good approximation) the Zipf law \cite{ZIP,KBLMS,OTT} (the Zipf 
law is the Pareto law with the exponent $\alpha=1$). In order to derive the Zipf law from Eq. (\ref{rown5}), 
we have to provide, therefore, functions $A(m)$ and $B(m)$ in the threshold form: 
\begin{eqnarray}
A(m) = \left\{ \begin{array}{ll}
A^<(m)=A_0+a\, m, & \textrm{if $m<m_1$} \\
{A^{\ge }(m)=A'_0}+a'\, m, & \textrm{if $m\ge m_1$}
\label{rown12}
\end{array} \right.
\end{eqnarray}
\begin{eqnarray}
B(m) = \left\{ \begin{array}{ll}
B^<(m)=B_0+b\, m^2 = b\, (m^2_0+m^2) & \\ \textrm{\hspace*{1.0cm} if $m<m_1$} \\
B^{\ge }(m)=B'_0+b'm^2 = b'(m'^2_0+m^2) & \\ \textrm{\hspace*{1.0cm} if $m\ge m_1$}
\label{rown13}
\end{array} \right.
\end{eqnarray}
where $m^2_0=B_0/b$ and $m'^2_0=B'_0/b'$. The threshold parameter $m_1$ can be interpreted as 
a crossover income between the medium- and high-income society classes. Remarkably, both income crossovers $m_0$ and 
$m_1(\ge m_0)$ are exogenous parameters. They should be determined from the dependence of the empirical complementary 
cumulative distribution function on variable $m$ because both crossovers are sufficiently distinct (see below for 
details).

Apparently, we assumed above that the formalism of the income change is the same for 
the whole society. This formalism is expressed by the threshold nonlinear Langevin equation where particular 
dynamics distinguishes the range of the high-income society class from those of the others.

For protection of the equilibrium distribution function against discontinuity at the threshold $m_1$ (which means 
adoption of the continuity principle of the equilibrium distribution function of household incomes being a kind of 
Ockham's razor principle), the following requirement should be satisfied:
\begin{eqnarray}
P_{\rm eq}^< (m=m_1)=P_{\rm eq}^{\ge }(m=m_1)
\label{rown14}
\end{eqnarray}
where
\begin{eqnarray}
P_{\rm eq}^< (m)&=&\frac{const}{B^<(m)}\exp\left(-\int_{m_{\rm init}}^{m(<m_1)}\frac{A^<(m')}{B^<(m')}\, dm'\right),
\label{rown14a}
\end{eqnarray}
and
\begin{eqnarray}
P_{\rm eq}^{\ge }(m)&=&\frac{const}{B^{\ge}(m)}\exp\left(-\int_{m_{\rm init}}^{m(\ge m_1)}\frac{A(m')}{B(m')}\, 
dm'\right) 
\nonumber \\
&=&\frac{const}{B^{\ge}(m)}\exp\left(-\int_{m_{\rm init}}^{m_1}\frac{A^<(m')}{B^<(m')}\, dm'\right) \nonumber \\
&\times &\exp\left(-\int_{m_1}^{m(\ge m_1)}\frac{A^{\ge }(m')}{B^{\ge }(m')}\, dm'\right)       .
\label{rown14b}
\end{eqnarray}
By substituting Eqs. (\ref{rown14a}) and (\ref{rown14b}) into Eq. (\ref{rown14}), we directly obtain
\begin{eqnarray}
B^<(m=m_1)&=&B^{\ge }(m=m_1) \nonumber \\
&\Leftrightarrow &B_0+b\, m^2_1=B'_0+b'm^2_1. 
\label{rown15b}
\end{eqnarray}
To assure that the interpretation of the parameter $m'_0$ is consistent with the income crossover $m_0$, we further put
\begin{eqnarray}
m'_0=m_0\Leftrightarrow \frac{B_0}{b}=\frac{B'_0}{b'}.
\label{rown16}
\end{eqnarray}
Moreover, in accordance with Eq. (\ref{rown16}), we make even more rigorous assumptions
\begin{eqnarray}
B'_0=B_0\; \; \mbox{and} \; \; b'=b.
\label{rown18}
\end{eqnarray}  

Subsequently, by substituting Eqs. (\ref{rown12}) and (\ref{rown13}) into Eqs. (\ref{rown14a}) and (\ref{rown14b}), we 
finally get
\begin{eqnarray}
P_{\rm eq}(m) =\left\{ \begin{array}{ll}
c'\, \frac{\exp\left(-(m_0/T)\arctan(m/m_0)\right)}{[1+(m/m_0)^2]^{(\alpha +1)/2}}, & \textrm{if $m<m_1$} \\
c''\, \frac{\exp\left(-(m_0/T_1)\arctan(m/m_0)\right)}{[1+(m/m_0)^2]^{(\alpha_1 +1)/2}}, & \textrm{if $m\ge m_1$}
\end{array} \right.
\label{rown19}
\end{eqnarray}
where $\alpha _1=1+a'/b$ and $T_1=B_0/A'_0$ while $c'$ and $c''$ are mutually related constants. These constants are
proportional to the normalisation factor $const$. Besides, constant $c'$ depends on $m_{\rm init},\, m_0,\, T$, and 
$\alpha $ while $c''$ additionally depends on $T_1,\, m_1$, and $\alpha _1$. Apparently, the number of free (effective) 
parameters driving the two-branch distribution function, Eq.  (\ref{rown19}), is reduced because this function depends only on 
the ratio of the initial parameters defining the Langevin dynamics (\ref{rown1}). 

For $m_1\gg m_0$, the interpretation of the distribution function, Eq. (\ref{rown19}), is self-consistent, as required, 
because the two power-law regimes are well defined. Then, for instance for $m\gg m_0$, the second branch in 
Eq. (\ref{rown19}) becomes the power-law dependence driven by the Pareto exponent $\alpha _1$ different (in general) from 
$\alpha$.

Importantly, our analysis indicates that the existence of the third income region is already allowed by theory. We are 
following this indication below.    

\section{Results and discussion}
In principle, we are ready to compare the theoretical complementary cumulative distribution function based on our 
probability distribution function $P_{\rm eq}(m)$, given by Eq. (\ref{rown19}), with the empirical data for the whole 
income range. However, the analytical form of this theoretical complementary cumulative distribution function is 
unknown in the closed explicit form. Therefore, we calculate it numerically. The key technical question arises on how to fit this complicated 
theoretical function to the empirical data. The fitting procedure consists of three steps as, fortunately, all 
parameters are to be found (in principle) by using independent fitting routines, as follows.

In the initial step, we found rough (more or less) approximated values of crossovers $m_0$ and $m_1$ directly from the 
plot of the empirical complementary cumulative distribution function (or empirical data). Thus, uncertainty of the 
$m_0$ and $m_1$ parameters did not exceed $10\%$, which was sufficiently accurate. Moreover, we took the exact value 
of the parameter $m_{\rm init}$ as the first point in the record of the empirical data. 

Secondly, we determined the temperature $T$ value by fitting the Boltzmann-Gibbs formula, Eq. (\ref{rown7}), to the 
corresponding empirical data in the range extending from $m_{\rm init}$ to $m_0$ (both found in the initial step). 
Notably, we assumed that this formula could be characterised by a single temperature value since the society as a whole 
was considered to be in (partial) equilibrium during the whole fiscal year. That is, we further put 
$T_1=T\Leftrightarrow A_0'=A_0$.

At the third step, we determined exponents $\alpha $ and $\alpha_1$ by separately fitting the weak Pareto law 
to the empirical data for the medium- and high-income society classes, respectively.
 
\begin{figure}[ht]
\centering
\includegraphics[scale=0.45]{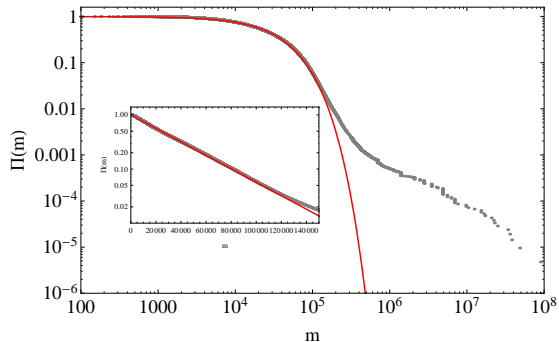}

\caption{\label{fig4} Fit of the (exponential) Boltzmann-Gibbs law, Eq. (\ref{rown7}), (solid line) to empirical data (dots) 
of the EU low-income society class for year 2008 \cite{EURO2, Forbes} in the log-log scale, for parameters 
$m_{\rm init}= 0.01$ EUR, $m_0=1.40\times 10^5\pm 0.14\times 10^5$ EUR and $T=34902 \pm 3$. The inset shows the fit in 
the log-linear scale.} 
\end{figure}

\begin{figure}[ht]
\centering
\includegraphics[scale=0.45]{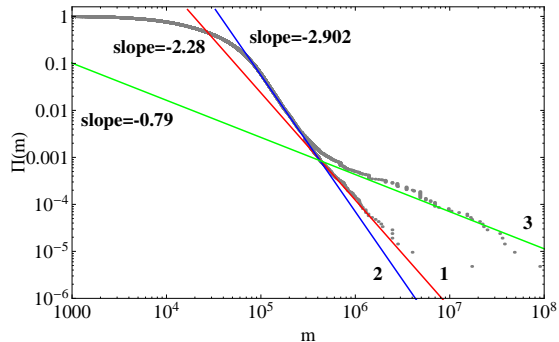}
\caption{\label{fig1} Curves 1 and 2: the fit of the weak Pareto law, Eq. (\ref{rown9}), (solid line) to empirical data (dots) of the EU 
medium-income society class for year 2008, for parameters $m_0=1.40\times 10^5\pm 0.14\times 10^5$ EUR, 
$m_1=4.0\times 10^5\pm 0.4\times 10^5$ EUR  and $\alpha =2.902 \pm 0.002$ if the Forbes database is included (curve 2) or $\alpha =2.28 \pm 0.01$ if the Forbes database is not included (curve 1). Curve 3: the 
fit of the weak Pareto law, Eq. (\ref{rown9}), to empirical data (dots) of the EU high-income society class 
for year 2008, for parameter $\alpha _1=0.79 \pm 0.01$ \cite{EURO2, Forbes}.}
\end{figure}

The results obtained in these three steps are correspondingly presented in Figs. \ref{fig4} and \ref{fig1}, mainly in 
the log-log scale (only in Fig. \ref{fig4} the inserted plot is presented in the log-linear scale).

The plot in Fig. \ref{fig4} shows the complementary cumulative exponential distribution function, i.e. the 
Boltzmann-Gibbs law, Eq. (\ref{rown7}), which quite well describes the EU low-income society class. This finding
significantly supports universal applicability of the Boltzmann-Gibbs law in economy.

Subsequently, the plot in Fig. \ref{fig1} was constructed. It quite well describes the EU medium-income society class 
by the weak Pareto law. Apparently, by joining the Forbes empirical database concerning
an effective income of billionaires with the EU-SILC database, we found that the Pareto (effective, nonuniversal) 
exponent increased from $\alpha =2.28 \pm 0.01$ to $\alpha =2.902 \pm 0.002$. This result defines the range of  Pareto 
exponents. This range covers, e.g. the exponent $\alpha = 2.67$ obtained very recently for the medium-income society 
class in Romania  for 2008 by considering a voluminous social security database \cite{arxiv}. However, this database 
contains only empirical data for low- and medium-income society class (in our terminology). In principle, it would 
be also possible to join this voluminous database with the Forbes corresponding dataset if Romania billionaires are 
present as members of the Forbes rank. As a result of our joining, the range of the medium-income society class became 
much narrower and shifted to incomes reduced by one order of magnitude. The medium-income society class is so sensitive 
to the size of the high-income society class as the former contains only no more than $3\%$ of all households.

These results are significant as they demonstrate a crucial role of two income society classes in the society structure, that 
is the low- and the high-income society classes.

Remarkably, we fitted again the weak Pareto law (\ref{rown9}) to the so completed empirical data by taking into account 
only the high-income society class (see plot in Fig. \ref{fig1}). Again, this class is well described by the weak Pareto
law, however, driven by the exponent $\alpha _1$ slightly lower than $1.0$ (cf. caption to Fig. \ref{fig1}). This 
result was expected, as the high-income households belonging to the high-income society class are usually the owners 
of companies, whose profits are described, indeed, by the Zipf law. 

Importantly, a power-law distribution reveals special property for exponent $\alpha _1\le 1$. That is, the first and 
higher moments (here moments of income) diverge. This means that the proper approach should apply quantiles (which 
are always finite \cite{HF}) instead of expectation values. For instance, the median should be used instead of the mean 
value. Anyway, the moment estimates are always finite and can be calculated directly from the empirical data. Moreover, 
there is no characteristic (physical) scale in the high-income society class as the mean value diverges. That is, the 
(hierarchical, self-similar in a probabilistic sense) income structure of the high-income society class is scale free 
making all levels of the structure equivalent. In other words, the same dynamical rules apply across the entire 
high-income society class independently of the particular income of different households (belonging to the high-income 
society class) \cite{KBLMS}.

\begin{figure}[ht]
\centering
\includegraphics[scale=0.45]{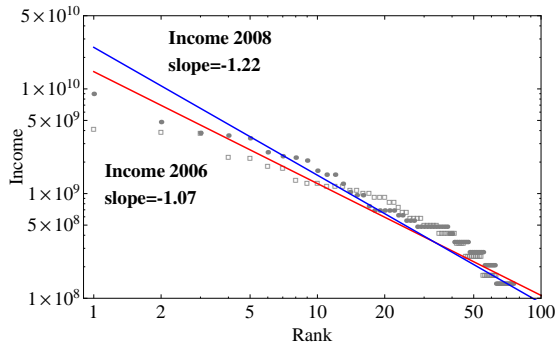}
\caption{\label{fig6} Ranks of incomes of billionaires in the EU. Solid lines were obtained by fitting straight lines (in 
the log-log scale) to empirical data (dots) for year 2008 
($\alpha_{\rm Pareto}=1/\alpha_{\rm rank}=1/1.22=0.82 \pm 0.02$) and to empirical data (squares) for year 2006 
($\alpha_{\rm Pareto}=1/\alpha_{\rm rank}=1/1.07=0.93 \pm 0.04$)
\cite{Forbes}.}
\end{figure}

To complete our analysis, we calculated the Pareto exponent for the high-income society class by using an alternative 
approach. We consider the rank of the difference between the wealth of billionaires in successive years (here year 2008 
and 2007 as well as 2006 and 2005). These ranks, i.e. straight lines in the log-log scale, are plotted in 
Fig. \ref{fig6}. The slopes of these lines (equal to $-\alpha_{\rm rank}$) were calculated using a fitting routine. 
The inverse of $\alpha_{\rm rank}$ gives the Pareto exponent $\alpha_{\rm Pareto}$.

\begin{figure}[ht]
\centering
\includegraphics[scale=0.45]{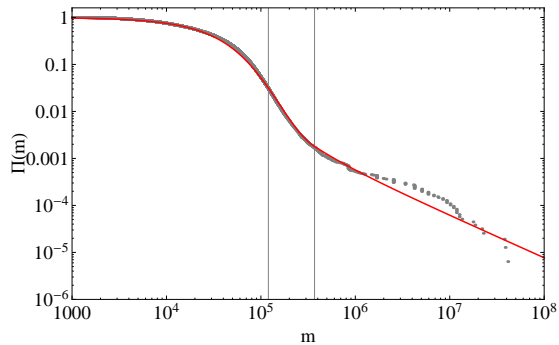}
\caption{\label{fig2} Fit of the complementary cumulative distribution function, based on the extended Yakovenko et al. 
formula, Eq. (\ref{rown19}), (solid line), to the EU household income empirical data 
(dots) for year 2006 ($T_1=T_2=T=43\times 10^3\pm 1\times 10^3$ EUR, $m_0=1.20\times 10^5\pm 0.12\times 10^5$ EUR, 
$m_1=3.70\times 10^5\pm 0.37\times 10^5$ EUR, $\alpha = 3.171\pm 0.002$ , $\alpha_1 = 0.90\pm 0.04$ ). The first and 
the second vertical line represents $m_0$ and $m_1$ \cite{EURO1, Forbes}, respectively.}
\end{figure}

Apparently, our calculation of Pareto exponent by using two independent methods gives almost the same 
results $(\alpha _1=0.79\pm 0.01$ and $\alpha _{Pareto}=0.82\pm 0.02)$ for year 2008, as expected \cite{ME,KBLMS}, suggesting that 
these methods are mutually consistent. We show below that the analogous conclusion is fulfilled also for year 
2006 (cf. Figs. \ref{fig6} and \ref{fig2}) as exponents $\alpha _1$ and $\alpha _{Pareto}$ are almost equal 
($\alpha _1 = 0.90\pm 0.04$ and $\alpha _{Pareto}=0.93\pm 0.04$).

\begin{figure}[ht]
\centering
\includegraphics[scale=0.45]{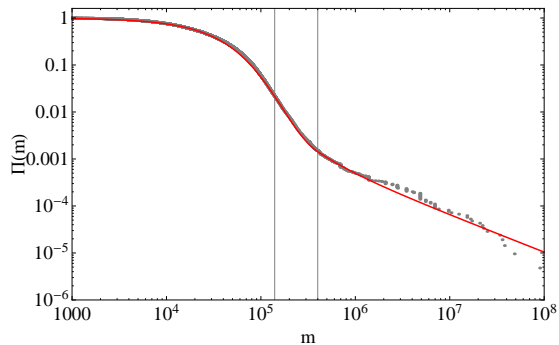}
\caption{\label{fig5} Fit of the complementary cumulative distribution function, based on the extended Yakovenko et al. 
formula (\ref{rown19}) (solid line), to the EU household income empirical data set (dots) for year 2008 
($T_1=T_2=T=39.5\times 10^3\pm 1\times 10^3$ EUR, $m_0=1.40\times 10^5\pm 0.14\times 10^5$ EUR 
$m_1=4\times 10^5\pm 0.4\times 10^5$ EUR, $\alpha = 2.902 \pm 0.002$, $\alpha_1 = 0.79 \pm 0.01$). The first and the
second vertical line represents $m_0$ and $m_1$ \cite{EURO2, Forbes}, respectively.}
\end{figure}

Hence, we have already obtained all values required by the extended Yakovenko et al. formula, Eq. (\ref{rown19}). 
The corresponding plots of the empirical and theoretical complementary cumulative distribution functions in the log-log 
scale are compared in Figs. \ref{fig2} and \ref{fig5} for years 2006 and 2008, respectively. Apparently, the 
predictions of the extended Yakovenko et al. formula, Eq. (\ref{rown19}), (solid curves in Figs. \ref{fig2} and \ref{fig5}) 
well agree with the empirical data (dots in Figs. \ref{fig2} and 
\ref{fig5})\footnote{The value of the income temperature $T$, obtained from the fit of the Boltzmann-Gibbs law to the 
empirical data points for the low-income society class, is $35321(\pm 4)$ EUR for  year 2006. However, the fit of the 
complementary cumulative distribution function, based on the Yakovenko et al. formula, Eq. (\ref{rown19}), to all empirical 
data points is slightly improved, as we used the income temperature $T$ higher by ca. $20\%$. }$^{,}$\footnote{For year 
2008, the value of the income temperature $T$, obtained from the fit of the Boltzmann-Gibbs law to the empirical data 
points for the low-income society class, is $34902(\pm 3)$ EUR. However, the fit of the complementary cumulative 
distribution function, based on the Yakovenko et al. formula, Eq. (\ref{rown19}), to all empirical data points is slightly 
improved, as we used the income temperature $T$ higher by ca. $10\%$.}.
That is, the extended Yakovenko et al. model well describes the empirical complementary cumulative distribution 
functions of the household incomes in the EU for all society classes, i.e. for the low-, medium-, and high-income society 
class. These successful descriptions  result from the sufficiently realistic assumptions of the model adopted herein. 
These assumptions allow for coexistence of additive and multiplicative processes as well as for the continuity 
principle of the equilibrium distribution function of the household incomes to be obeyed. 

\section{Conclusions}
Herein, we proved that the household incomes of all society classes in the EU can be modelled by the 
nonlinear threshold Langevin dynamics (\ref{rown1}) with $m$-dependent drift term, $A(m)$, and $m$-dependent dispersion, 
$B(m)$, given by Eq. (\ref{rown12}) and (\ref{rown13}), respectively. At the threshold $m_1$, there is a jump of the 
proportionality coefficient of the drift term. That is, this term abruptly changes from $a$ to $a'$, where $a'<a$ (as 
$\alpha _1<\alpha $). It means that the stochastic term in Eq. (\ref{rown1}) is relatively more significant in this 
case (i.e. the above threshold $m_1$) than the drift term. That is, economic activity of the high-income society class is much 
more risky than activities of all other society classes, as expected.

By comparing results obtained for years 2006 and 2008 (cf. Figs. \ref{fig2} and \ref{fig5}), we found that the threshold
$m_1$ was only slightly higher for the latter year than for the former. It means that  in year 2006, to the high-income society class belonged the society members almost as rich as those belonging to this class in year 2008. Moreover,
the result of only slightly more extensive society stratification in year 2008 was confirmed by exponents' inequality, 
as the exponent $\alpha _1$ for year 2006 is only slightly higher than that for year 2008. Furthermore, the 
slowly-varying tendency (similar to that considered above) is also observed for the medium-income society class. In 
fact, it is surprising how stable with respect to the recent financial crisis the shape of the curve $\Pi(m)$ vs. $m$ 
is. That is, only the number of households belonging to a given income society class most likely changed but the income 
structure of the society as a whole was not altered. 

The completed database, which we used (by properly joining the Forbes empirical database with that of EU-SILC), 
emphasises a significant role of the high-income society class. Namely, the presence of the third region increases this 
Pareto exponent which characterises the medium-income society class making the range of this class narrower and shifting 
it to incomes lower by one order of magnitude (cf. Fig. \ref{fig1}). This latter society class is now so much reduced 
that it occupies almost an intermediate region between low- and high-income society classes. Apparently, the role of the 
low- and medium-income society classes was, in the present case, significantly reduced.

The use of two different datasets (EU-SILC and Forbes), which are not necessarily compatible, carries a methodological 
danger of getting discontinuous fits. Nevertheless, bringing these two datasets together and analysing them jointly is 
better for making progress in this field then avoiding their comparison.

Herein, we succeeded in comparing ratios (i.e. relative population of successive income society classes) of 
$r_1=\frac{\Pi (m_{init})-\Pi (m_0)}{\Pi (m_0)-\Pi (m_1)}$ and $r_2=\frac{\Pi (m_0)-\Pi (m_1)}{\Pi (m_1)}$ for both 
year 2006 and 2008 by using our formula, Eq. (\ref{rown19}). Hence, we determined $r_1=32.66$ and $r_2=16.48$ for year 2006 
as well as $r_1=48.98$ and $r_2=13.97$ for year 2008. We obtained information on, relatively, how many society members 
belong to a given income society class. Apparently, population of the medium-income society class is strongly decreased in year 2008 
in comparison to that in year 2006. Several members of this class were shifted both to the low- and to high-income society 
classes. Our low-parameter approach seems to be much more sensitive than that using the Gini coefficient (G) \cite{Gini} because we obtained $G=54.34$ and $G=54.89$ for year 2006 and 2008, respectively. 

Furthermore, we estimated the percentage  breakdown of population of the society classes: for year 2006 -- low-income: 
$96.85\%$; medium-income: $2.97\%$; high-income: $0.18\%$ and for year 2008 -- low-income: $97.86\%$; medium-income: 
$2.00\%$; high-income: $0.14\%$. These results can be considered as complementary to that (obtained above) corresponding
to the relative population of successive income society classes. Interestingly, the total fraction of the medium- and 
high-income classes in the EU was around $3\%$ in year 2006, which is about the same as that found by Yakovenko et al. 
\cite{Yako2} for the US, and this fraction has decreased to around $2\%$ in year 2008, most likely due to the financial 
crisis.

Economists often argue that economic and political conditions are quite different in the US and in the EU, and expect a lower 
income inequality in the EU. However, we demonstrate quantitatively herein that the exponential law does apply to the EU as well.  This finding gives much stronger support for universal applicability of the Boltzmann-Gibbs law in economics.  

Our work shows that the income distribution in the low-income class, covering around 97\% of population, follows the 
exponential law, whereas in the two upper-income classes the distribution follows two power laws with different 
exponents. Remarkably, the role of the medium-income 
society class is strongly reduced making it an intermediate one within our approach to the complementary cumulative distribution function.

\section*{Acknowledgements}
We thank Victor M. Yakovenko and Tiziana Di Matteo for very stimulating comments and suggestions.

\end{document}